\documentclass[aps,prb,twocolumn,showpacs,superscriptaddress,groupedaddress]{revtex4}
\usepackage{graphicx}
\usepackage{epstopdf}
\usepackage{dcolumn}
\usepackage{bm}
\usepackage{amssymb}
\usepackage{amsmath}

\begin{document}

\title{Orbital Dependent Electronic Masses in Ce Heavy Fermion Materials
studied via Gutzwiller Density Functional\ Theory}
\date{\today}
\author{Ruanchen Dong}
\affiliation{Department of Physics, University of California Davis, Davis, California
95616, USA}
\author{Xiangang Wan}
\affiliation{National Laboratory of Solid State Microstructures and Department of
Physics, Nanjing University, Nanjing 210093, China}
\author{Xi Dai}
\affiliation{Beijing National Laboratory for Condensed Matter Physics and Institute of
Physics, Chinese Academy of Sciences, Beijing 100190, China}
\author{Sergey Y.~Savrasov}
\affiliation{Department of Physics, University of California Davis, Davis, California
95616, USA}

\begin{abstract}
A series of Cerium based heavy fermion materials is studied using a
combination of local density functional theory and many--body Gutzwiller
approximation. Computed orbial dependent electronic mass enhancements
parameters are compared with available data extracted from measured values
of Sommerfeld coefficient. The Gutzwiller density functional theory is shown
to remarkably follow the trends across a variety of Ce compounds, and to
give important insights on orbital selective mass renormalizations that
allow better understanding of wide spread of data.
\end{abstract}

\pacs{}
\maketitle

\section{Introduction}

Heavy fermion materials pose one of the greatest challenges in condensed
matter physics. Their low--temperature linear specific heat coefficient can
be up to 1000 times larger than the value expected from the free--electron
theory; their magnetic moments can be screened by the Kondo effect and their
electrical resistivity is frequently divergent but sometimes
superconductivity can emerge at low temperatures \cite{hewson}.

Theoretical calculations based on density functional theory (DFT) in its
popular Local Density Approximation (LDA) \cite{LDA} fail to reproduce
strongly renormalized electronic masses in heavy fermion materials due to
improper treatment of many--body correlation effects. Consider, for example,
a well--known class Cerium class of heavy fermion materials, such as two
famous phases ($\alpha $ and $\gamma $) of Cerium itself \cite%
{panousis1970,johansson1974,allen1982}, so called Ce--115s systems: CeXIn$%
_{5}$ (X=Co,Rh,Ir) \cite{petrovic2001,hegger2000,petrovic2001_2} and
numerous Ce--122 compounds \cite%
{liang1988,mihalik,nakamoto2004,lu2005,steglich1979,hunt1990,harima1991,steglich1985,thompson1985,besnus1985,graf1997,dhar1987,steeman1988}%
: CeX$_{2}$Si$_{2}$ (X=Mn,Fe,Co,Ni,Cu,Ru,Rh,Pd,Ag). A Table 1 gives
evaluated via LDA\ densities of states specific heat coefficients $\gamma $
for these materials as compared to experiments, where in many cases a factor
of 2--20 error exists in underestimating $\gamma $ while in some cases,
such, e.g., as CeCo$_{2}$Si$_{2}$, an overestimation occurs. It is the
purpose of this work to show that a better treatment of electronic
correlations via a recently introduced Gutzwiller Density Functional Theory 
\cite{ho2008,deng2008,LDA+G,yao2011} can correct most of these errors and
uncover which exactly orbitals become heavy as is also illustrated in Table
I.

The physics of the electronic mass enhancement is controlled by a low
frequency behavior of the local electronic self--energy which can be encoded
in a simple Teilor like form%
\begin{equation}
\Sigma _{\alpha }(\omega )=\Sigma _{\alpha }(0)+(z_{\alpha }^{-1}-1)\omega
+...  \label{Sigma}
\end{equation}%
where we assume, for simplicity, the existence of some crystal field
representation $|\alpha \rangle $ that diagonalizes the self--energy matrix
in a spin--orbital space of the localized f electrons. This expression
suggests two main effects that may occur when correlations are brought into
consideration on top of a band theory such as LDA: first the crystal field
correction to a local f--electron level is controlled by $\Sigma _{\alpha
}(0)$ and, second, the actual band narrowing is controlled by a
quasiparticle residue parameter $z_{\alpha }.$ Both effects would affect our
comparisons of $\gamma $ in Table 1.

Recently, advanced many body approaches based on combinations of Density
Functional\ and Dynamical Mean Field Theory (DMFT) have been implemented 
\cite{LDA+DMFT} to study heavy fermion systems \cite{shim2007,MunehisaPRL}
where self--consistent solutions of either Anderson or Kondo impurity
problems have been done using most accurate Continues Time Quantum Monte
Carlo method \cite{CTQMC-Werner, CTQMC-Junya}. While DMFT deals with full
frequency dependent self--energy and is a lot more computationally demanding
than traditional LDA, it taught us an important lesson on the so called
orbital selectivity in the Mott transition problem, i.e. when crystal field
dependent self--energies can reduce effective degeneracy of the impurity.
This affects the proximity of the quasiparticle residue $z$ to become equal
zero when the ratio between Hubbard $U$ and bandwidth $W$ changes.

A recently introduced Gutzwiller Density Functional Theory (GDFT) and the so
called LDA+G method \cite{ho2008,deng2008,LDA+G,yao2011} is a simplified
variational approach that relies on the Gutzwiller approximation initially
introduced to study itinerant ferromagnetism of the one--band Hubbard model\ 
\cite{gutzwiller}, and later extended to multi--band systems \cite%
{bunemann1997,bunemann1998}. In this method, the atomic configurations of
correlated orbitals are treated by adjusting their weights using a
variational procedure. This leads to renormalized energy bands and mass
enhancements for the electrons. The approximation was extensively studied in
the limit of infinite dimensions~\cite{metzner1989,metzner1989_2}, and was
shown to be equivalent to a slave--boson mean--field theory~\cite%
{kotliar1986} for both single--band and multiband models~\cite%
{gebhard1991,lechermann2007,bunemann2007}. In LDA+G, the Gutzwiller type
trial wave function $\hat{P}|0\rangle $ is adopted with $|0\rangle $ and $%
\hat{P}$ being the LDA ground state and the Gutzwiller projector
respectively. After further application of the Gutzwiller approximation, an
effective Hamiltonian describing the dynamics of quasiparticles was obtained
as $H_{eff}=\hat{P}H_{LDA}\hat{P}$, which contains two important features
discussed above: the modification of the crystal/spin--orbital fields and
the quasiparticle weight $z_{\alpha }$. Thus, the method exactly casts the
effect encoded into the low--frequency behavior of $\Sigma (\omega )$, Eq.
(1).

The benchmark of the LDA+G scheme was demonstrated in Ref.~%
\onlinecite{deng2008}. For a non--magnetic correlated metal $\mathrm{SrVO}%
_{3}$, it produced narrower bands and larger effective masses than those
found in standard LDA. Also the method was able to get a photoemission peak
missed in the LDA calculation. These improvements are very close to the
experimental results. Later, in Ref.~\onlinecite{wang2008} a complex phase
diagram of $\mathrm{Na}_{x}\mathrm{CoO}_{3}$ was correctly reproduced.
Recently, this method has been successfully applied to FeAs--based
superconductors~\cite{wang2010,yao2011} and to Ce metal \cite%
{tian2011,KotliarCe}.

\begin{table*}[tbp]
\caption{Comparison between calculated using LDA density of states and
experimentally extracted specific heat coefficients $\protect\gamma $ and
the extracted quasiparticle residues $z_{\exp }=\protect\gamma _{LDA}/%
\protect\gamma _{\exp }$ for a number of Cerium based heavy fermion
compounds considered in this work. The last columns show the predictions of $%
\protect\gamma $ and $z$ using the LDA+G method with the values of U=4 and 5
eV as well as the reference to a specific orbital degeneracy of the $j=5/2$
manifold that exibits strongest enhancement.}
\label{table1}%
\begin{ruledtabular}
\begin{tabular}{lccccrccccl}
                               & $N_{LDA}(0)$            & $\gamma _{LDA}$                & $\gamma _{exp}$      & Ref.                                              & $z_{\exp }$   & $ \gamma _{LDA+G}$ & $ \gamma _{LDA+G}$    & $ z_{LDA+G}$ & $ z_{LDA+G}$ & Orbital                             \\ 
                               & St./(Ry$\cdot $cell)    & mJ/(mol $K^{2}$)               & mJ/(mol $K^{2}$)     & to $\gamma _{exp}$                                &               &  (U=4 eV)          &  (U=5 eV)             &  (U=4 eV)    &  (U=5 eV)    & $j=5/2$                             \\ \hline
$\alpha $-Ce                   & 36                      & 6.2                            & 13                   & \onlinecite{panousis1970}                         & 0.48          &                    &                       &   0.55       & 0.33         & $\Gamma _{7}$,$\Gamma _{8}$         \\ 
$\gamma $-Ce                   & 49                      & 8.5                            &                      &                                                   &               &                    &                       &   0.14       & 0.11         & $\Gamma _{7}$,$\Gamma _{8}$         \\ \hline
115s:                          &                         &                                &                      &                                                   &               &                    &                       &              &              &                                     \\ 
$\text{CeCoIn}_{5}$            & 150                     & 26.0                           & 290                  & \onlinecite{petrovic2001}                         & 0.09          &  70                & 104                   &   0.20       & 0.13         & $\Gamma _{6}$,$2\times \Gamma _{7}$ \\ 
$\text{CeRhIn}_{5}$            & 156                     & 27.2                           & 420                  & \onlinecite{hegger2000}                           & 0.065         &  75                & 120                   &   0.19       & 0.12         & $\Gamma _{6}$,$2\times \Gamma _{7}$ \\ 
$\text{CeIrIn}_{5}$            & 165                     & 28.6                           & 720                  & \onlinecite{petrovic2001_2}                       & 0.04          &  79                & 210                   &   0.14       & 0.07         & $\Gamma _{6}$,$2\times \Gamma _{7}$ \\ \hline
122s:                          &                         &                                &                      &                                                   &               &                    &                       &              &              &                         \\ 
$\text{CeMn}_{2}\text{Si}_{2}$ & 184                     & 31.9                           & 47                   & \onlinecite{liang1988}                            & 0.68          &  52                & 84                    &   0.51       & 0.43         & $\Gamma _{6}$,$2\times \Gamma _{7}$ \\ 
$\text{CeFe}_{2}\text{Si}_{2}$ & 85                      & 14.7                           & 22                   & \onlinecite{mihalik}                              & 0.67          &  21                & 24                    &   0.53       & 0.47         & $\Gamma _{6}$,$2\times \Gamma _{7}$ \\ 
$\text{CeCo}_{2}\text{Si}_{2}$ & 110                     & 19.0                           & 10                   & \onlinecite{nakamoto2004}                         & 1.9?          &  35                & 43                    &   0.47       & 0.36         & $\Gamma _{6}$,$2\times \Gamma _{7}$ \\ 
$\text{CeNi}_{2}\text{Si}_{2}$ & 115                     & 19.9                           & 33                   & \onlinecite{lu2005}                               & 0.60          &  27                & 29                    &   0.47       & 0.43         & $\Gamma _{6}$,$2\times \Gamma _{7}$ \\ 
$\text{CeCu}_{2}\text{Si}_{2}$ & 64                      & 11.1                           & 1000                 & \onlinecite{steglich1979,hunt1990,harima1991}     & 0.01          &  430               & $\infty $             &   0.10       & ~0           & $\Gamma _{6}$                       \\ 
$\text{CeRu}_{2}\text{Si}_{2}$ & 103                     & 17.8                           & 350                  & \onlinecite{steglich1985,thompson1985,besnus1985} & 0.05          &  70                & 190                   &   0.33       & 0.13         & $2\times \Gamma _{7}$               \\ 
$\text{CeRh}_{2}\text{Si}_{2}$ & 106                     & 18.3                           & 130                  & \onlinecite{graf1997}                             & 0.14          &  35                & 120                   &   0.36       & 0.14         & $2\times \Gamma _{7}$               \\
$\text{CePd}_{2}\text{Si}_{2}$ & 100                     & 17.3                           & 65-110               & \onlinecite{dhar1987,steeman1988}                 & 0.15-0.26     &  170               & $\infty $             &   0.045      & ~0           & $\Gamma _{7}$                       \\ 
$\text{CeAg}_{2}\text{Si}_{2}$ & 205                     & 35.5                           &                      &                                                   &               &  140               & 430                   &   0.10       & 0.06         & $\Gamma _{6}$
\end{tabular}
\end{ruledtabular}
\end{table*}

In this work we address the physics of Cerium heavy fermion materials via
the use of the LDA+G approach. A great spread in the extracted values of
mass enhancements data shown in Table I together with some unphysical values
of $z>1$ prompts us that in many classes of real compounds both the orbital
selectivity encoded via the shifts $\Sigma _{\alpha }(0)$ on top of the LDA\
as well as the quasiparticle residues play an important role and have to be
treated on the same footing. It therefore represents a stringiest test of a
many body electronic structure method such as LDA+G to heavy fermion
materials. In particular, in its recent application to elemental Cerium \cite%
{KotliarCe}, it has been shown that the spin--orbital splitting of the
f--level is renormalized by correlations and pushes energies of the J=7/2
manifold up relative to J=5/2 states$.$ This resulted in lowering the
degeneracy from 14 to 6 and in a greater mass enhancement of J=5/2 manifold
as compared to a non--spin orbit coupled calculation.

The paper is organized as follows. In Sec. II, the LDA+G method is
described. The results for several typical families of heavy fermion
materials are presented in Sec. III. Finally, Sec. IV is the conclusion.

\section{The LDA+G Method}

The Gutzwiller Density Functional Theory and the LDA+G approximation have
been described previously \cite{ho2008,deng2008,LDA+G,yao2011}. Here we
merely summarize the equations of the method that we implement using a
linear muffin--tin orbital formalism that includes both the full potential
terms and relatvistic spin--orbit coupling operator variationally \cite%
{FPLMTO} .

\subsection{Gutzwiller Approximation}

We first illustrate the method using a general multi--orbital Hubbard model.
The Hamiltonian is: 
\begin{equation}
H=H_{0}+H_{int}=\sum_{ij,\alpha \beta }t_{ij}^{\alpha \beta }c_{i\alpha
}^{\dagger }c_{j\beta }+\sum_{i}\sum_{\alpha \neq \beta }U_{i}^{\alpha \beta
}\hat{n}_{i\alpha }\hat{n}_{i\beta }  \label{HUB}
\end{equation}%
where $\alpha =1,\ldots ,2N$ is the spin--orbital index of the localized
orbital, $N$ is the number of orbitals, e.g. 7 for the $f$--orbital. The
first term is a tight--binding Hamiltonian which can be extracted from the
LDA calculation. The second term is the on--site interaction which has been
restricted to the density--density interaction.

In the atomic limit, for the localized orbital there are $2N$ different
states which can be either occupied or empty. Therefore there is total $%
2^{2N}$ configurations $|\Gamma \rangle $. All these configurations form a
complete basis and the density--density interaction is diagonal in this
configuration space. It is obvious that these configurations should not be
equally weighted. In the Gutzwiller method, we adjust the weight of each
configuration. Therefore it is convenient to construct projection operators
that project onto a specific configuration $\Gamma $ at site $i$: 
\begin{equation}
\hat{m}_{i\Gamma }=|i\Gamma \rangle \langle i\Gamma |
\end{equation}

When the interaction term is absent, the ground state is the Hatree
uncorrelated wave function (HWF) $|\Psi _{0}\rangle $ which is a Slater
determinant of the single--particle states. When the interaction is switched
on, this wave function is no longer a good approximation. In the Gutzwiller
method, we project the wave function to a Gutzwiller wave function (GWF) $%
|\Psi _{G}\rangle $ by adjusting the weight of each configuration through
variational parameters $\lambda _{i\Gamma }$ ($0\leqslant \lambda _{i\Gamma
}\leqslant 1$): 
\begin{equation}
|\Psi _{G}\rangle =\hat{\mathcal{P}}|\Psi _{0}\rangle =\prod_{i}\hat{P}%
_{i}|\Psi _{0}\rangle
\end{equation}%
where 
\begin{equation}
\hat{P}_{i}=\sum_{\Gamma }\lambda _{i\Gamma }\hat{m}_{i\Gamma }
\end{equation}%
Notice that when all $\lambda _{i\Gamma }=1,$ the GWF is going back to HWF.
At the same time, setting $\lambda _{i\Gamma }=0$ removes configuration $%
\Gamma $ at site $i$. Therefore, perfectly localized atomic state of site $i$
is described by all $\lambda _{i\Gamma }=0$ except for one, and in this way,
the Gutzwiller wave function captures both the itinerant and localized
behavior of the system.

It is a difficult task to evaluate GWF. However, within the Gutzwiller
method, we can map any operator $\hat{O}$ acting on the GWF to a
corresponding effective $\hat{A}^{G}$ which acts on the HWF: 
\begin{equation}
\langle \Psi _{G}|\hat{A}|\Psi _{G}\rangle =\langle \Psi _{0}|{\hat{\mathcal{%
P}}}^{\dag }\hat{A}\hat{\mathcal{P}}|\Psi _{0}\rangle =\langle \Psi _{0}|%
\hat{A}^{G}|\Psi _{0}\rangle
\end{equation}%
where 
\begin{equation}
\hat{A}^{G}={\hat{\mathcal{P}}}^{\dag }\hat{A}\hat{\mathcal{P}}
\end{equation}%
Specifically, when the operator $\hat{A}$ is a single--particle operator,
e.g. $\hat{A}=\sum_{ij,\alpha \beta }A_{ij}^{\alpha \beta }c_{i\alpha
}^{\dag }c_{j\beta }$ where $A_{ij}^{\alpha \beta }=\langle i\alpha |\hat{A}%
|j\beta \rangle $, the Gutzwiller effective operator can be written as: 
\begin{equation}
\hat{A}^{G}=\sum_{ij,\alpha \beta }\sqrt{z_{i\alpha }}A_{ij}^{\alpha \beta }%
\sqrt{z_{j\beta }}c_{i\alpha }^{\dag }c_{j\beta }+\sum_{i,\alpha
}A_{ii}^{\alpha \alpha }(1-z_{i\alpha })c_{i\alpha }^{\dag }c_{i\alpha }
\end{equation}%
where $z_{i\sigma }$ are the orbital--dependent quasiparticle residues: $%
0\leqslant z_{i\alpha }\leqslant 1.$ These are determined by the
configuration weights: 
\begin{equation}
z_{i\alpha }=\sum_{\Gamma \Gamma ^{\prime }}\frac{\sqrt{m_{i\Gamma
}m_{i\Gamma ^{\prime }}}|\langle i\Gamma ^{\prime }|c_{i\alpha }^{\dag
}|i\Gamma \rangle |}{\sqrt{n_{i\alpha }(1-n_{i\alpha })}}
\end{equation}%
where $m_{i\Gamma }=\langle \Psi _{G}|\hat{m}_{i\Gamma }|\Psi _{G}\rangle $
and $n_{i\alpha }$ are the occupation numbers for the orbitals.

\begin{widetext}

\subsection{Combination with LDA}

Similar to the idea of the LDA+U or LDA+DMFT methods, we add the interaction
term on top of the LDA calculation. The Hamiltonian is given by: 
\begin{equation}
H=H_{LDA}+H_{int}-H_{DC}
\end{equation}%
where $H_{\mathrm{LDA}}$ is the LDA Hamiltonian, which casts the same form
as $H_{0}$ in Eq.(\ref{HUB}), $H_{\mathrm{int}}$ is the on--site interaction
term for the set of correlated orbitals, such as $f$--orbitals of
heavy--fermion materials considered in this work. Since the LDA calculation
has already included the Coulomb interaction in some averaged level, we need
to subtract the double--counting term $H_{DC}$ from LDA. Various forms of $%
H_{DC}$ will be discussed later.

The Kohn--Sham approach uses a non--interacting system as a reference which
keeps the same density as the interacting one. However, now our ground state
is the GWF instead of the HWF. Therefore, we need to transform the
Hamiltonian into the effective one in the $|\Psi _{0}\rangle $ basis. Since
the Hamiltonian $H_{LDA}$ is a single--particle operator, following Ref.\ %
\onlinecite{LDA+G} we obtain

\begin{eqnarray*}
H_{LDA}^{G} &=&\langle \Psi _{G}|H_{LDA}|\Psi _{G}\rangle = \\
&&\left( \sum_{\alpha i}\sqrt{z_{\alpha }}|\phi _{\alpha i}\rangle \langle
\phi _{\alpha i}|+1-\sum_{\alpha i}|\phi _{\alpha i}\rangle \langle \phi
_{\alpha i}|\right) H_{LDA}\left( \sum_{\beta j}\sqrt{z_{\beta }}|\phi
_{\beta j}\rangle \langle \phi _{\beta j}|+1-\sum_{\beta j}|\phi _{\beta
j}\rangle \langle \phi _{\beta j}|\right) + \\
&&\sum_{\alpha i}(1-z_{\alpha })|\phi _{\alpha i}\rangle \langle \phi
_{\alpha i}|H_{LDA}|\phi _{\alpha i}\rangle \langle \phi _{\alpha i}|
\end{eqnarray*}%
where $|\phi _{\alpha i}\rangle $represents a complete basis set of the
correlated orbitals, and where we omit site index $i$ from the quasiparticle
residues $z_{a}$due to lattice periodicity.

The interaction term acting on the GWF produces: 
\begin{equation}
H_{int}^{G}=\langle \Psi _{G}|H_{int}|\Psi _{G}\rangle =\sum_{i\Gamma
}E_{\Gamma }m_{\Gamma }
\end{equation}

The expectation value of the total Hamiltonian gives us the total energy as
a functional of the density $\rho $ and confugurational weights $m_{\Gamma }$%
: 
\begin{equation}
E(\rho ,\{m_{\Gamma }\})=\langle \Psi _{0}|H_{LDA}^{G}|\Psi _{0}\rangle
+\sum_{\Gamma }E_{\Gamma }m_{\Gamma }-E_{DC}
\end{equation}

A minimization similar to LDA is now performed. Representing density in
terms of the Kohn--Sham states, produces the equations for the
quasiparticles:

\begin{equation}
\frac{\partial E(\rho ,\{m_{\Gamma }\})}{\partial \langle \psi _{nk}}=\left(
H_{LDA}^{G}+\sum_{\alpha }\left[ \frac{\partial E}{\partial z_{\alpha }}%
\frac{\partial z_{\alpha }}{\partial n_{\alpha }}-\frac{\partial E_{DC}}{%
\partial n_{\alpha }}\right] |\phi _{\alpha }\rangle \langle \phi _{\alpha
}|\right) |\psi _{nk}\rangle =\epsilon _{nk}|\psi _{nk}\rangle  \label{Heff}
\end{equation}

\begin{equation}
\frac{\partial E(\rho )}{\partial m_{\Gamma }}=\sum_{\alpha }\frac{\partial E%
}{\partial z_{\alpha }}\frac{\partial z_{\alpha }}{\partial m_{\Gamma }}%
+E_{\Gamma }=0
\end{equation}%
Recalling the self--energy linear expansion, Eq.(\ref{Sigma}), we see from
Eq.(\ref{Heff}) that the effective Hamiltonian to be diagonalized casts the
following form 
\begin{equation}
H^{G}=H_{LDA}^{G}+\sum_{\alpha }(\Sigma _{\alpha }(0)-V_{DC,\alpha
})z_{\alpha }|\phi _{\alpha }\rangle \langle \phi _{\alpha }|
\end{equation}%
where $\Sigma _{\alpha }(0)$ and $V_{DC,\alpha }$ are directily associated
with various total energy derivatives appeared in (\ref{Heff}).

\subsection{Gutzwiller Projected Hamiltonian}

It is convinient to represent all matrices in the space of the Bloch
eigenvalues $\epsilon _{\mathbf{k}j}$ and wave functions $|\mathbf{k}%
j\rangle $ that are obtained from the LDA calculation. The Gutzwiller
hamiltonian to be diagonalized is given by

\begin{eqnarray}
\langle \mathbf{k}j^{\prime }|H^{G}|\mathbf{k}j\rangle &=&\sum_{j^{\prime
\prime }}\epsilon _{\mathbf{k}^{\prime \prime }j^{\prime \prime }}\left(
\sum_{\alpha }\sqrt{z_{\alpha }}\langle \mathbf{k}j^{\prime }|\phi _{\alpha
}\rangle \langle \phi _{\alpha }|\mathbf{k}j^{\prime \prime }\rangle +\delta
_{j^{\prime }j^{\prime \prime }}-\sum_{\alpha }\langle \mathbf{k}j^{\prime
}|\phi _{\alpha }\rangle \langle \phi _{\alpha }|\mathbf{k}j^{\prime \prime
}\rangle \right) \times  \label{ham} \\
&&\left( \sum_{\beta }\sqrt{z_{b}}\langle \mathbf{k}j^{\prime \prime }|\phi
_{\beta }\rangle \langle \phi _{\beta }|\mathbf{k}j\rangle +\delta
_{j^{\prime \prime }j}-\sum_{\beta }\langle \mathbf{k}j^{\prime \prime
}|\phi _{\beta }\rangle \langle \phi _{\beta }|\mathbf{k}j\rangle \right)
+\sum_{\alpha }(1-z_{\alpha })\langle \mathbf{k}j^{\prime }|\phi _{\alpha
}\rangle \epsilon _{\alpha }\langle \phi _{\alpha }|\mathbf{k}j\rangle + 
\notag \\
&&\sum_{\alpha }\left( \Sigma _{\alpha }(0)-V_{DC,\alpha }\right) z_{\alpha
}\langle \mathbf{k}j^{\prime }|\phi _{\alpha }\rangle \langle \phi _{\alpha
}|\mathbf{k}j\rangle  \notag
\end{eqnarray}%
where a subset of correlated orbitals $|\phi _{\alpha }\rangle $ is
introduced. Their levels are given by%
\begin{equation}
\epsilon _{\alpha }=\sum_{\mathbf{k}^{\prime \prime }j^{\prime \prime
}}\epsilon _{\mathbf{k}^{\prime \prime }j^{\prime \prime }}\langle \phi
_{\alpha }|\mathbf{k}^{\prime \prime }j^{\prime \prime }\rangle \langle 
\mathbf{k}^{\prime \prime }j^{\prime \prime }|\phi _{\alpha }\rangle
\end{equation}%
The quasiparticle residues $z_{\alpha }$ and the level shifts $\Sigma
_{\alpha }(0)$ are obtained using the Gutzwiller procedure \cite{LDA+G}. The
double counting potential $V_{DC,a}$ corrects for the fact that the LDA
already accounts for some of the correlation effects in a mean field manner.
The eigenvalue problem

\begin{equation}
\sum_{j}(\langle \mathbf{k}j^{\prime }|H^{G}|\mathbf{k}j\rangle -\delta
_{j^{\prime }j}E_{\mathbf{k}n})B_{j}^{\mathbf{k}n}=0
\end{equation}%
produces renormalized energy bands $E_{\mathbf{k}n}$ and wave functions $%
\sum_{j}B_{j}^{\mathbf{k}n}|\mathbf{k}j\rangle $ of the quasiparticles$.$

\subsection{Charge Density}

In order to find a new density, we calculate a Gutzwiller density matrix
operator in the LDA representation

\begin{eqnarray}
\rho _{j^{\prime }j}^{G\mathbf{k}} &=&\sum_{j^{\prime \prime }}f_{\mathbf{k}%
j^{\prime \prime }}\left( \sum_{\alpha }\sqrt{z_{\alpha }}\langle \mathbf{k}%
j^{\prime }|\phi _{\alpha }\rangle \langle \phi _{\alpha }|\mathbf{k}%
^{\prime \prime }\rangle +\langle \mathbf{k}j^{\prime }|\mathbf{k}j^{\prime
\prime }\rangle -\sum_{\alpha }\langle \mathbf{k}j^{\prime }|\phi _{\alpha
}\rangle \langle \phi _{\alpha }|\mathbf{k}j^{\prime \prime }\rangle \right)
\times \\
&&\left( \sum_{\beta }\sqrt{z_{\beta }}\langle \mathbf{k}j^{\prime \prime
}|\phi _{\beta }\rangle \langle \phi _{\beta }|\mathbf{k}j\rangle +\langle 
\mathbf{k}j^{\prime \prime }|\mathbf{k}j\rangle -\sum_{\beta }\langle 
\mathbf{k}j^{\prime \prime }|\phi _{\beta }\rangle \langle \phi _{\beta }|%
\mathbf{k}j\rangle \right) +\sum_{\alpha }\langle \mathbf{k}j^{\prime }|\phi
_{\alpha }\rangle (1-z_{\alpha })\rho _{\alpha }\langle \phi _{a}|\mathbf{k}%
j\rangle  \notag
\end{eqnarray}%
where 
\begin{equation}
\rho _{\alpha }=\sum_{\mathbf{k}j}f_{\mathbf{k}j}\langle \mathbf{k}j|\phi
_{\alpha }\rangle \langle \phi _{\alpha }|\mathbf{k}j\rangle
\end{equation}%
Diagonalizing it produces new occupation numbers $n_{\mathbf{k}\lambda }$ 
\begin{equation}
\sum_{j}(\rho _{j^{\prime }j}^{G\mathbf{k}}-\delta _{j^{\prime }j}n_{\mathbf{%
k}\lambda })C_{j}^{\mathbf{k}\lambda }=0
\end{equation}%
so that the density of quasiparticles in real space is given by%
\begin{equation}
\rho ^{G}(\mathbf{r})=\sum_{\mathbf{k}\lambda }n_{\mathbf{k}\lambda }\left(
\sum_{j^{\prime }}C_{j^{\prime }}^{\mathbf{k}\lambda }|\mathbf{k}j^{\prime
}\rangle \right) \left( \sum_{j}C_{j}^{\mathbf{k}\lambda \ast }\langle 
\mathbf{k}j|\right)
\end{equation}

\end{widetext}

\subsection{Incompleteness of Basis}

To see the importance of the issue, let us examine a shift of the LDA
eigenstates $\epsilon _{\mathbf{k}j}$ by arbitrary value $x$. We obtain the
new Gutzwiller Hamiltonian as the old one plus the correction%
\begin{equation}
\langle \mathbf{k}j^{\prime }|\tilde{H}^{G}|\mathbf{k}j\rangle =\langle 
\mathbf{k}j^{\prime }|H^{G}|\mathbf{k}j\rangle +xo_{j^{\prime }j}(\mathbf{k})
\end{equation}%
where $o_{j^{\prime }j}(\mathbf{k})$ is a matrix that can be proved to be
equal to $\delta _{j^{\prime }j}$ only under the assumption that the LDA
wave functions form a mathematically complete basis set, i.e.%
\begin{equation}
\sum_{j^{\prime \prime }}\langle \phi _{\alpha }|\mathbf{k}j^{\prime \prime
}\rangle \langle \mathbf{k}j^{\prime \prime }|\phi _{\beta }\rangle =\delta
_{\alpha \beta }
\end{equation}%
Unfortunately, modern electronic structure methods deal with finite basis
sets, and the last relationship is only approximately satisfied. As a
result, different choices of energy zero for the LDA eigenvalues $\epsilon _{%
\mathbf{k}j}$ may lead to slightly different output, although in our
application to well--localized Cerium 4f electrons, this introduces only
minor noise in our calculated results. In the following, we always assume
that the LDA eigenvalues are measured with respect to the Fermi energy which
is the only physically relevant energy in this problem.

\subsection{Double Counting Potential}

As one sees from Eq.(\ref{ham}) the actual self--energy correction used in
the LDA+G calculation is $\Sigma _{\alpha }(0)-V_{DC,\alpha }.$ Frequently,
a so called LDA+U version \cite{LDA+U} of double counting potential $%
V_{DC,\alpha }$ is used, that for our case is just an orbital--independent
energy shift given by 
\begin{equation}
V_{DC}^{LDA+U}=U(n_{f}-1/2),
\end{equation}%
where $n_{f}$ is the average number of f electrons which at Ce f--shell is
close to unity. As one sees, this correction has just an overall level shift
by $U/2$ and does not modify the Gutzwiller extracted spin--orbit and
crystal fields encoded in the $\alpha $ dependence of $\Sigma _{\alpha }(0).$
Unfortunately, it is not exactly clear whether the overall level shift of Ce
f electrons has a physical effect, since the standard LDA+U double counting
was introduced in connection to the Hartree Fock value of the self--energy
which is the value at infinite frequency, $\Sigma (\infty )$. It therefore
may not be suited for correcting low energy physics of the heavy fermion
systems. 

As a result, in this work we adopt a different strategy in order to
elucidate the physics of orbital selectivity in Ce heavy fermion compounds:
our calculations are first performed without $\Sigma _{a}(0)$ correction
assuming that the double counting potential 
\begin{equation}
V_{DC,\alpha }^{(1)}=\Sigma _{\alpha }(0).
\end{equation}%
It has an important justification that the LDA\ calculated Fermi surfaces do
not acquire any modifications as was the past evidence for some heavy
fermion uranium compounds\cite{Norman}. Second, we introduce the crystal
field averaged double counting 
\begin{equation}
V_{DC}^{(2)}=\frac{1}{N}\sum_{\alpha }^{N}\Sigma _{\alpha }(0)
\end{equation}%
which keeps the average position of the f--level intact but allows for its
crystal field modifications found self--consistently via the LDA+G
procedure. Since both $V_{DC,\alpha }^{(1)}$ and $V_{DC,\alpha }^{(2)}$ rely
on Gutzwiller extracted $\Sigma _{\alpha }(0),$ which itself is obtained
from the LDA+G functional minimization procedure, the entire method is still
variational and allows an accurate estimate of the total energy. Comparing
calculations with two types of double counting, important conclusions can be
drawn on which exactly orbitals of a given heavy fermion system play a major
role in its electronic mass enhancement.

\section{Results and Discussion}

We are interested in calculating the mass enhancement parameters of Cerium f
electrons which in a simple single band theory would be given by the ratio
of $m^{\ast }/m_{LDA}$. A common approach to extract this data is to compare
the values of the Sommerfeld coefficient $\gamma $ evaluated using the LDA\
density of states at the Fermi level $N_{LDA}(0)$%
\begin{equation}
\gamma _{LDA}=\frac{\pi ^{2}}{3}k_{B}N_{LDA}(0)
\end{equation}%
with the measured electronic specific heat which, according to the Fermi
liquid theory, behaves at low temperatures as $\gamma T$. However, some care
should be taken when adopting this procedure. First, densities of states of
real systems include muliband features and contributions from both heavy and
light electrons. Second, LDA densities of states assume some crystal field
effects which should in general be supplemented by many body corrections
encoded in $\Sigma _{a}(0).$ Therefore not only band narrowing but also
level shifts are expected to occur in real life on top of LDA. Third, many
of the materials discussed in our work undergo either antiferromagnetic or
superconducting transition before reaching $T\rightarrow 0$ limit. We quote
the data for $\gamma _{\exp }$ in Table I using the data for their lowest
temperature paramagnetic phases.

There are several other ways to access this information that we are going to
use in this work. Optical spectroscopy experiments can provide access to
effective masses but those are frequency dependent. Angle--resolved
photoemission spectroscopy (ARPES) experiments measure directly
one--electron spectral functions 
\begin{equation}
A(\mathbf{k},\omega )=\frac{|\Im \Sigma (\mathbf{k},\omega )|}{(\omega -\Re
\Sigma (\mathbf{k},\omega ))^{2}+\Im \Sigma ^{2}(\mathbf{k},\omega )}
\end{equation}%
Here $\Re \Sigma $ is the real part and $\Im \Sigma $ is the imaginary part
of the electronic self--energy. Under the assumption of the locality of
self--energy, the quasiparticle residues 
\begin{equation}
z=\left( 1-\frac{\partial \Re \Sigma (\omega )}{\partial \omega }\right)
^{-1}
\end{equation}%
can be extracted by comparing ARPES spectra against calculated LDA energy
bands.

The dHvA effect is another powerful experimental technique which measures
the properties of the Fermi surface properties under applied magnetic field~%
\cite{onuki2012}. LDA calculation can identify each cyclotron orbit seen by
the dHvA experiment and find corresponding effective masses. However,
complexity in shapes of 3D Fermi surfaces in real systems also makes this
method not perfect.

It is remarkable that the LDA+G calculation returns the orbital dependent
quasiparticle residues $z_{\alpha }$ directly. In Table I, we are quoting
these mass enhancement data and \emph{not} the ones obtained via our
calculated LDA+G densities of states.

\subsection{$\protect\alpha$-Ce and $\protect\gamma$-Ce}

\begin{figure}[tbp]
\includegraphics[width=\columnwidth]{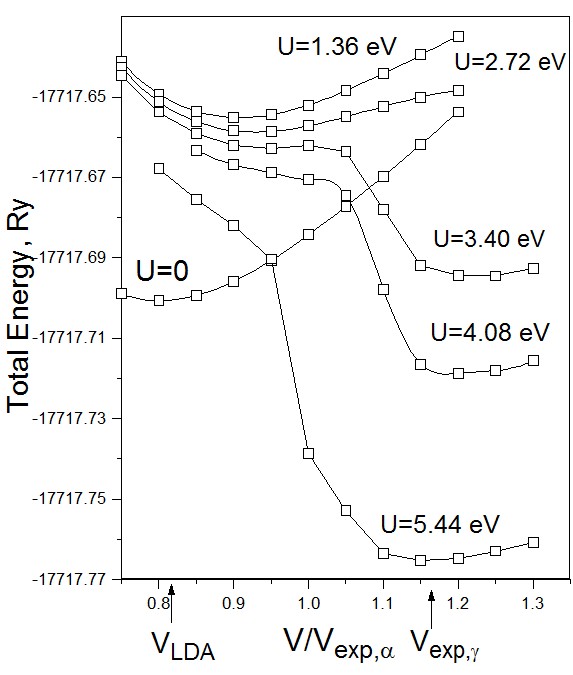}
\caption{(color online) Calculated total energy vs volume using the LDA+G
method with several values of Hubbard U. }
\label{CeEtot}
\end{figure}

We first discuss our calculations for Cerium metal which is famous for its
iso--structural phase transition from its $\alpha $ to $\gamma $ phase that
is accompanied by 15\% of its volume expansion, and has attracted great
attention in the past\cite%
{johansson1974,allen1982,zolfl2001,held2001,haule2005,amadon2006} and current%
\cite{KotliarCe} literature. It is remarkable that our LDA+G calculation can
correct most of the error in predicting the volume of the $\alpha $ phase
for both types of the double countings that we explore in this work.
Moreover, as Fig. \ref{CeEtot} illustrates, it clearly shows a double well
type of behavior of the energy vs volume (with smaller/larger minima
corresponding to $\alpha /\gamma $ phases) when using the crystal field
averaged double counting $V_{DC}^{(2)}$ and the Hubbard U's within the range
between 3.5 and 5.5 eV. This is in accord with the previous previous
LDA+DMFT studies for Cerium \cite{held2001}. Similar behavior has been also
found when studying $\alpha \rightarrow \delta $ transition in metallic
Plutonium \cite{PuNature}.

The specific heat measurement of $\alpha $--Ce gives its Sommerfeld
coefficient $\gamma \sim 13$ mJ/(mol$\cdot \mathrm{K}^{2}$).~\cite%
{panousis1970}. The optical spectroscopy experiment~\cite{vandereb2001}
estimates the effective mass to be $\sim 6m_{e}$ in $\alpha $--Ce and to be $%
\sim 20m_{e}$ in $\gamma $--Ce, indicating the itinerant/localized features
of the $\alpha /\gamma $ phases. However, the estimated optical effective
masses are frequency dependent.

\begin{figure}[tbp]
\includegraphics[width=\columnwidth]{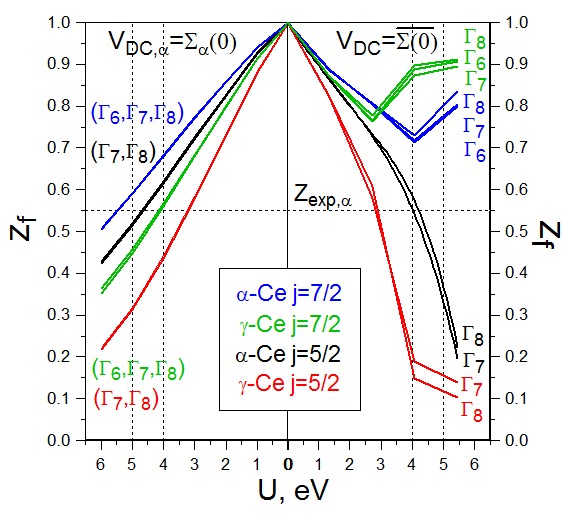}
\caption{(color online) Calculated dependence of quasiparticle residues as a
function of Hubbard U in $\protect\alpha $ and $\protect\gamma $ phases of
Cerium metal using the LDA+G approach.}
\label{Cerium}
\end{figure}

Fig. \ref{Cerium} shows the dependence of the quasiparticle residues as a
function of the Hubbard U for volumes corresponding to $\alpha $-- and $%
\gamma $--Ce where the left/right plots represent our calculations with $%
V_{DC}^{(1)}/V_{DC}^{(2)}$ type of double countings. It is clear when
crystal/spin--orbital corrections are not taken into account (left plot),
the effective masses for various Cerium orbitals are very similar in values
and are not very strongly enhanced even for large values of U. (We use
relativistic cubic harmonics representation where $j=5/2$ level is split
onto $\Gamma _{7}$, $\Gamma _{8}$ and $j=7/2$ onto $\Gamma _{6},\Gamma
_{7},\Gamma _{8}$ states.) The situation changes dramatically when we
account for the local self--energy correction (right plot): a Coulomb
renormalized spin--orbit splitting pushes $\Gamma _{7},\Gamma _{8}$ states
of $j=5/2$ manifold down and $\Gamma _{6},\Gamma _{7},\Gamma _{8}$ states of 
$j=7/2$ manifold up. This results in redistributing the occupancies of the
f--electrons which now reside mainly at $j=5/2$ level. Thus, the effective
degeneracy is 6 instead of 14 and the quasiparticle residues become much
more sensitive to the values of U.

Actual comparison of the quasiparticle residues with experiment needs an
accurate estimate of the Hubbard U for Cerium f--electrons which is
typically around 5 eV, although, if the Hunds rule $J$\symbol{126}1eV
(important for the\ virtual f$^{2}$ state) is taken into account, the
effective interaction is reduced a little bit to $U-J.$ Due to a great
sensitivity of quasiparticle residues to this parameter, and for the
purposes of our work, we simply show in Table I the range of $z$'s that one
can obtain for the values of U in the range between 4 and 5 eV. For $\alpha $%
--Ce we find them between 0.55 and 0.33 which are close to their
experimental estimates. This is also consistent with the result of the
LDA+DMFT calculation~\cite{sakai2005}. While there is no specific heat data
for the $\gamma $ phase, optical measurements \cite{vandereb2001} show a
factor of 3--4 enhanced masses as compared to the ones of the $\alpha $ phase%
$,$ and our reduced values of $z^{\prime }$s are in accord with this trend.

\subsection{Ce-115s}

We next discuss our applications to the so--called 115 series of Cerium
heavy fermion compounds $\text{CeMIn}_{5}$ ($\text{M}$ = Co, Rh, Ir). They
have a tetragonal $\text{HoCoGa}_{5}$--type structure which results in
additional splitting of all $\Gamma _{8}$ quadruplets into $\Gamma _{6}$ and 
$\Gamma _{7}$ doublets. Despite having similar structure and almost
identical LDA electronic structures, these systems show very different
properties which has attracted a great interest. $\text{CeCoIn}_{5}$ is a
heavy fermion superconductor with critical temperature $T_{c}=2.3$ K,
highest known in Ce--based systems\ \cite{petrovic2001} and with the
Sommerfeld coefficient $\gamma \cong 290$ mJ/(mol$\cdot \text{K}^{2}$)
measured just above T$_{c}$. $\text{CeIrIn}_{5}$ is also a superconductor
with T$_{c}=0.4$ K \cite{petrovic2001_2} with $\gamma $=720 mJ/(mol$\cdot 
\mathrm{K}^{2}$) above T$_{c}$ that is nearly temperature independent. $%
\text{CeRhIn}_{5}$, on the other hand, is an antiferromagnet with N\'{e}el
temperature $T_{N}$=3.8 K but becomes a superconductor with $T_{c}$= 2.1 K
above a critical pressure $P_{c}\sim $16 kbar \cite{hegger2000}. From the $%
C/T$ data, there is a peak at $T_{N}$=3.8 K, indicating the onset of
magnetic ordering. In order to find the electronic specific heat, one needs
to use isostructural, nonmagnetic $\mathrm{LaRhIn}_{5}$ to subtract the
lattice contribution to $C$. However, it is difficult to define precisely
the electronic specific heat above $T_{N}$ due to the peaked structure. A
simple entropy--balance construction gives a Sommerfeld coefficient $\gamma
\geqslant 420$ mJ/(mol$\cdot \mathrm{K}^{2}$) for $T>T_{N}$.

Those different properties are considered as the result of the localized vs
itinerant nature of the 4$f$ electrons. The dHvA measurements for $\text{%
CeCoIn}_{5}$ have shown the effective cyclotron masses within the range from
9 to 20 $m_{0}$ which is consistent with the specific heat data~\cite%
{hall2001, settai2001}. The LDA calculation with a model of itinerant $f$
electrons shows a reasonable agreement with the dHvA data~\cite{elgazzar2004}
while complete localization of the $f$ electrons is needed to get the
agreement with the angle--resolved photoemission spectroscopy(ARPES) data 
\cite{koitzsch2009}. The dHvA experiment has been also performed for the
antiferromagnetic state of $\text{CeRhIn}_{5}$ \cite{cornelius2000,
hall2001_2}. Although earlier LDA calculation with the itinerant model shows
some agreement with the experimental data~\cite{hall2001_2}, the localized
nature of the $f$ electrons was confirmed by the dHvA measurements~in $\text{%
Ce}_{x}\text{La}_{1-x}\text{RhIn}_{5}$ \cite{alver2001} and by comparing the
Fermi surfaces between $\text{CeRhIn}_{5}$ and $\text{LaRhIn}_{5}$~\cite%
{shishido2002}. With the application of pressure, the dramatic change of the
Fermi surface was observed indicating the change from the localized
antiferromagnetic state to the itinerant heavy fermion state \cite%
{shishido2005}. For $\text{CeIrIn}_{5}$, there is some experimental
controversy. The effective cyclone mass $m_{c}^{\ast }$ is observed in the
range from 6.3 to 45$m_{e}$ indicating a large enhancement \cite{haga2001}.
LDA\ with itinerant f electrons~\cite{haga2001,elgazzar2004} explains well
geometry and the volume of the Fermi surface but the band masses are much
smaller of the cyclotron masses. The photoemission spectrum is well
described by the LDA+DMFT calculation~\cite{shim2007} where the degree of
itineracy in $\text{CeIrIn}_{5}$ is though to be even larger than in $\text{%
CeCoIn}_{5}$ \cite{haule2010}. Also, this method shows the calculated
effective masses to be of the same order with the experimental one \cite%
{choi2012}. On the other hand, the ARPES study~\cite%
{fujimori2003,fujimori2006} shows that $\text{CeIrIn}_{5}$ and $\text{CeRhIn}%
_{5}$ have nearly localized 4$f$ electrons.

\begin{figure}[tbp]
\includegraphics[width=\columnwidth]{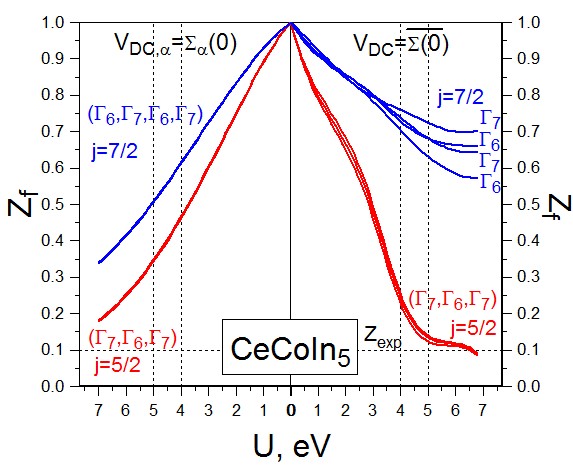}
\caption{(color online) Calculated dependence of quasiparticle residues as a
function of Hubbard U in CeCoIn$_{5}$ using the LDA+G approach.}
\label{CeCoIn5}
\end{figure}

The results of our paramagnetic LDA+G calculations for all three 115
compounds are presented in Table I. Similar to our calculation for Cerium,
we find that the calculated effective masses are only moderately enhanced ($z%
\symbol{126}0.3-0.5$) if we do not account for the crystal field/spin orbit
corrections to the self--energy on top of the LDA. When we perform
self--consistent calculation including the level shifts, much smaller values
of the quasiparticle residues can be reached. Fig. \ref{CeCoIn5} illustrates
this behavior for CeCoIn$_{5}.$ The situation here is similar to Cerium
where the spin--orbit coupling gets renormalized by correlations making the
effective degeneracy of the f--electrons equal 6. Actual values of the
quasiparticle residues in Table I are given for U equal 4 and 5 eV:\ we see
that the estimated z is the largest for $\text{CeCoIn}_{5}$ while the f
electrons are more localized in $\text{CeRhIn}_{5}$ and $\text{CeIrIn}_{5}$.
The residual discrepancies can be attributed to the above discussed
uncertainties seen in experiments and also to the intrinsic error connected
to the Gutzwiller procedure as our prior studies of the performance of this
method against Quantum Monte Carlo approach have shown a 30\% type of error%
\cite{InterpolativeSolver}. 

\subsection{Ce-122s}

We finally discuss our applications to Ce 122 types of systems. By itself, $%
RM_{2}X_{2}$ is an enormous class of ternary intermetallic compounds with
over hundreds of members, where $R$ is a rare--earth element, $M$ denotes a
transition metal (3$d$. 4$d$, or 5$d$) and $X$ is either silicon or
germanium. After its first discovery of superconductivity with a transition
temperature $T_{c}\simeq $0.5 K in $\text{CeCu}_{2}\text{Si}_{2}$~\cite%
{steglich1979}, the interest in this family awakens, especially due to the
interplay between antiferromagnetic and superconducting orders. Here we
focus on the subclass $\text{Ce}M_{2}\text{Si}_{2}$ ($M$ = $\text{Mn}$, $%
\text{Fe}$, $\text{Co}$, $\text{Ni}$, $\text{Cu}$, $\text{Ru}$, $\text{Rh}$, 
$\text{Pd}$, $\text{Ag}$) where all members have body--centered tetragonal $%
\text{ThCr}_{2}\text{Si}_{2}$--type structure with space group $I4/mmm$.

For $M=3d$ series, no magnetic order is found except for $\text{CeMn}_{2}%
\text{Si}_{2}$ where the Mn local moments order below 379 K \cite%
{szytula1981}. For $\text{CeCu}_{2}\text{Si}_{2}$, the electronic specific
heat coefficient $\gamma \simeq 1000$ mJ/(mol$\cdot \text{K}^{2}$) is the
largest one among this family. On the other hand, $\text{CeFe}_{2}\text{Si}%
_{2}$, $\text{CeCo}_{2}\text{Si}_{2}$ and $\text{CeNi}_{2}\text{Si}_{2}$ are
weak paramagnets with relatively small values of $\gamma $ value which is
shown in Table I. These three compounds are also known as valence
fluctuation systems.

For $M=4d$ series, first, $\text{CeRu}_{2}\text{Si}_{2}$ is known as a
archetypal Kondo lattice compound: it is a paramagnet with a relative large $%
\gamma \simeq 350$ mJ/(mol$\cdot \text{K}^{2}$). The other three are
antiferromagnets at low temperatures: $\text{CeRh}_{2}\text{Si}_{2}$ has the
highest ordering temperature 36--39 K~\cite{grier, quezel1984} while $\text{%
CePd}_{2}\text{Si}_{2}$~\cite{grier, steeman1988} and $\text{CeAg}_{2}\text{%
Si}_{2}$~\cite{grier, palstra} order antiferromagnetically at 8.5-10 K an
8-10 K, respectively. With the application of pressure, superconductivity
was found in $\text{CeRh}_{2}\text{Si}_{2}$~\cite{movshovich1996} and in $%
\text{CePd}_{2}\text{Si}_{2}$~\cite{mathur1998}. $\text{CeRu}_{2}\text{Si}%
_{2}$ is unique in this subclass since the superconductivity is not observed
down to a few mK. This makes it best target material for studying its heavy
fermion state. Interestingly, here a metamagnetic transition was found and
extensively studied~\cite{haen1987}. The cyclotron effective mass $m_{c}\sim
120m_{e}$ is observed in the dHvA experiment indicating the renormalized
heavy fermion state~\cite{aoki1992}. The electronic structure calculation
using LDA with itinerant $f$--electron model qualitatively explains the dHvA
data~\cite{yamagami1992,yamagami1993,runge1995} The metamagnetic transition
was also studied by dHvA experiments~\cite{onuki1992, aoki1993, aoki1993prl}
demonstrating that the $f$--electron character is changed from itinerant to
localized across the metamagnetic transition. All members of this family of
compounds were put into the Doniach ($T_{N}$ vs $J_{K})$ phase diagram~\cite%
{endstra1993}. Later, using the LDA+DMFT scheme, the phase diagram was
renewed~\cite{MunehisaPRL}.

\begin{figure}[tbp]
\includegraphics[width=\columnwidth]{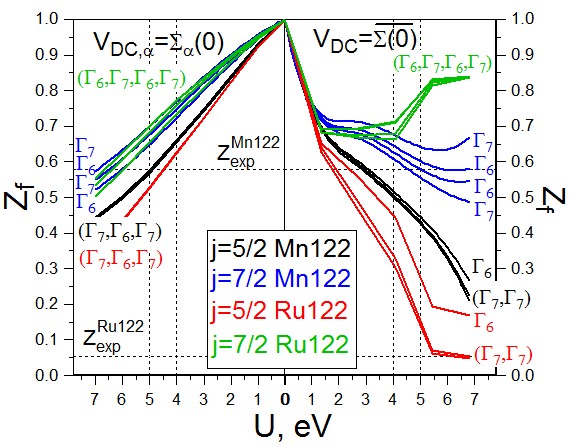}
\caption{(color online) Calculated dependence of quasiparticle residues as a
function of Hubbard U in CeMn$_{2}$Si$_{2}$ and CeRu$_{2}$Si$_{2}$ using the
LDA+G approach.}
\label{Ce122}
\end{figure}

Results of our applications to Ce 122s are presented in Table 1. They assume
a paramagnetic heavy fermion state for all systems and the experimental $%
\gamma ^{\prime }s$ are extracted from the specific heat data measured above
the temperatures of antiferromagnetic/superconducting transition. For the
122 systems with 3d elements such as Mn,Fe,Co,Ni, the LDA+G procedure
returns only moderately enhanced electron masses which we find in agreement
with the experiment. An example of the dependence of $z_{\alpha }$ vs $U$ is
shown on Fig. \ref{Ce122} for Mn based 122, where the Coulomb renormalizing
spin--orbit splitting is essential to reduce the orbital degeneracy from 14
to 6 when using the crystal field averaged double counting, $V_{DC}^{(2)}$.
This is similar to our findings in Ce and Ce--115 systems. Our calculation
for Cu based 122 shows that its quasiparticle residues are very sensitive to
the values of U above 4 eV. In fact, it is beginning to reach almost zero
values when U approaches 5 eV. It is also known experimentally that this
system shows enormous mass enhancement just before it goes into
superconducting state. 

The 122 compounds with 4d elements (Ru, Rh, Pd, Ag) exhibit strongly
renormalized quasiparticle masses. Our LDA+G calculations listed in Table I
correctly follow this trend where we see a strong reduction of z's as
compared to our calculations with 3d elements. We also find another
interesting effect: with increasing U the effective degeneracy of the f
electrons acquires a reduction not only due to the Coulomb assisted
renormalization of spin--orbit splitting but also renormalization of the
crystal fields: Fig. \ref{Ce122} shows this behavior for Ru-122 where we see
that the values of z become different for various crystal field levels of
the $j=5/2$ manifold: one $\Gamma _{6\text{ }}$and two $\Gamma _{7}$
doublets. We find that a similar effect occurs in all other 4d types of Ce
122s and the last column of Table I lists our results showing which
particular orbitals exhibit the strongest mass enhancement.

\section{Conclusion}

In conclusion, using a recently proposed LDA+G approach we have studied
quasiparticle mass renormalizations in several classes of Ce heavy fermion
compounds. We find that the calculation gives correct trends across various
systems as compared to the measured Sommerfeld coefficient and reproduces
the order of magnitude of the experimental value. We also uncover an
interesting orbital dependency of the quasiparticle residues for each
studied compound which provides an important physical insight on how
correlations affect the effective degeneracy of Cerium f--electrons placed
in various crystallographic environments.

\begin{center}
\textbf{Acknowledgements}
\end{center}

The authors are grateful to Y.--F. Yang for useful comments. This work was
supported by US DOE Nuclear Energy University Program under Contract No.
00088708.


\begin{thebibliography}{99}
\bibitem{hewson} For a review, see, e.g., A.~C.~Hewson, \emph{The Kondo
Problem to Heavy Fermions} (Cambridge University Press, Cambridge, 1997).

\bibitem{LDA} For a review, see, e.g., Theory of Inhomogeneous Electron Gas,
ed.by S. Lundqvist and N. H. March (Plenm, New York, 1983).

\bibitem{panousis1970} N.~T.~Panousis, K.~A.~Gschneidner Jr.,Solid State
Commun. \textbf{8}, 1779 (1970).

\bibitem{johansson1974} B.~Johansson, Philos. Magn. \textbf{30}, 469 (1974).

\bibitem{allen1982} J.~W.~Allen, R.~M.~Martin, Phys. Rev. Lett. \textbf{49},
1106 (1982).

\bibitem{petrovic2001} C.~Petrovic, P.~G.~Pagliuso, M.~F.~Hundley,
R.~Movshovich, J.~L.~Sarrao, J.~D.~Thompson, Z.~Fisk, and P.~Monthoux, J.
Phys.: Condens. Matter \textbf{13}, L337 (2001).

\bibitem{hegger2000} H.~Hegger, C.~Petrovic, E.~G.~Moshopoulou,
M.~F.~Hundley, J.~L.~Sarrao, Z.~Fisk, and J.~D.~Thompson, Phys. Rev. Lett. 
\textbf{84}, 4986 (2000).

\bibitem{petrovic2001_2} C.~Petrovic, R.~Movshovich, M.~Jaime,
P.~G.~Pagliuso, M.~F.~Hundley, J.~L.~Sarrao, Z.~Fisk, and J.~D.~Thompson,
Eur. Phys. Lett. \textbf{53}, 354. (2001).

% Mn gamma

\bibitem{liang1988} G.~Liang, I.~Perez, D.~DiMarzio, M.~Croft,
D.~C.~Johnston, N.~Anbalagan and T.~Mihalisin, Phys. Rev. B \textbf{37},
5970 (1988).

%Fe gamma

\bibitem{mihalik} M.~ Mihalik, M.~ Mihalik, and V.~Sechovsk\'{y}, Physica B 
\textbf{359--361}, 163 (2005).

%Co gamma

\bibitem{nakamoto2004} G.~Nakamoto, A.~Fuse, M.~Kurisu, and T.~Shigeoka, J.
Magn. Magn. Mater. \textbf{272-276}, e75 (2004).

%Ni gamma

\bibitem{lu2005} J.~J.~Lu, M.~K.~Lee, Y.~M.~Lu, J.~F.~Chou, L.Y. Jang, Solid
State Commun. \textbf{135}, 505 (2005).

%Cu gamma

\bibitem{steglich1979} F.~Steglich, J.~Aarts, C.~D.~Bredl, W.~Lieke,
D.~Meschede, W.~Franz, and H.~Sch$\ddot{a}$fer, Phys. Rev. Lett. \textbf{43}%
, 1892 (1979).

\bibitem{hunt1990} M.~Hunt, P.~Meeson, P.-A.~Probst, P.~Reinders,
M.~Springford, W.~Assmus and W.~Sun. J. Phys.: Condens. Matter \textbf{2},
6859 (1990).

\bibitem{harima1991} H.~Harima and A.~Yanase, J. Phys. Soc. Jpn. \textbf{60}%
, 21 (1991).

%%Ru gamma

\bibitem{steglich1985} F.~Steglich, U.~Rauchschwalbe, U.~Gottwick,
H.~M.~Mayer, G.~Sparn, N.~Grewe, U.~Poppe, and J.~J.~M.~Franse, J. Appl.
Phys. \textbf{57}, 3054 (1985).

\bibitem{thompson1985} J.~D.~Thompson, J.~O.~Willis, C.~Godart,
D.~E.~MacLaughlin, and L.~C.~Gupta, Solid State Commun. \textbf{56}, 169
(1985).

\bibitem{besnus1985} M.~J.~Besnus, J.~P.~Kappler, P.~Lehmann, and A.~Meyer,
Solid State Commun. \textbf{55}, 779 (1985).

%Rh gamma

\bibitem{graf1997} T.~Graf, J.~D.~Thompson, M.~F.~Hundley, R.~Movshovich,
Z.~Fisk, D.~Mandrus, R.~A.~Fisher, and N.~E.~Phillips, Phys. Rev. Lett. 
\textbf{78}, 3769 (1997).

%%%% Pd gamma

\bibitem{dhar1987} S.~K.~Dhar, E.~V.~Sampathkumaran, Phys. Lett. A \textbf{%
121}, 454 (1987).

\bibitem{steeman1988} R.~A.~Steeman, E.~Frikkee, R.~B.~Helmholdt,
A.~A.~Menovsky, J.~Van den Berg, G.~J~Nieuwenhuys, and J.~A.~Mydosh, Solid
State Commun. \textbf{66}, 103 (1988).

%%%%%%%%%%%%%%%%%%%%%%%%%%%%%%%

\bibitem{deng2008} X.~Y.~Deng, X.~Dai, and Z.~Fang, Eur. Phys. Lett. \textbf{%
83}, 3008 (2008).

\bibitem{ho2008} K.~M.~Ho, J.~Schmalian, and C.~Z.~Wang, Phys. Rev. B 
\textbf{77}, 073101 (2008).

\bibitem{LDA+G} X. Y. Deng, L. Wang, X. Dai, and Z. Fang, Phys. Rev. B 
\textbf{79}, 075114 (2009);

\bibitem{yao2011} Y.~X.~Yao, C.~Z.~Wang, and K.~M.~Ho, Phys. Rev. B \textbf{%
83}, 245139 (2011).

\bibitem{LDA+DMFT} For a review, see, e.g., G.~Kotliar, S.~Y.~Savrasov,
K.~Haule, V.~S.~Oudovenko, O.~Parcollet, and C.~A.~Marianetti, Rev. Mod.
Phys. \textbf{78}, 865 (2006).

\bibitem{shim2007} J.~H.~Shim, K.~Haule, and G.~Kotliar, Science \textbf{318}%
, 1615 (2007).

\bibitem{MunehisaPRL} M. Matsumoto, M. J. Han, J. Otsuki, S. Y. Savrasov,
Phys. Rev Lett. \textbf{103}, 096403 (2009).

\bibitem{CTQMC-Werner} P. Werner, A. Comanac, L. de' Medici, M. Troyer, and
A. J. Millis, Phys. Rev. Lett. \textbf{97}, 076405 (2006).

\bibitem{CTQMC-Junya} J. Otsuki, H. Kusunose, P. Werner, and Y. Kuramoto, J.
Phys. Soc. Jpn. \textbf{76}, 114707 (2007).

\bibitem{gutzwiller} M.~C.~Gutzwiller, Phys. Rev. Lett. \textbf{10}, 159
(1963); Phys. Rev. \textbf{134}, A923 (1964); \textbf{137}, A1726 (1965).

\bibitem{bunemann1997} J.~B\"{u}nemann, F.~Gebhard, and W.~Weber, J. Phys.:
Condens. Matter \textbf{9}, 7343 (1997).

\bibitem{bunemann1998} J.~B\"{u}nemann, W.~Weber, and F.~Gebhard, Phys. Rev.
B \textbf{57}, 6896 (1998).

\bibitem{metzner1989} W.~Metzner and D.~Vollhardt, Phys. Rev. Lett. \textbf{%
62}, 324 (1989).

\bibitem{metzner1989_2} W.~Metzner, Z. Phys. B: Condens. Matter \textbf{77},
253 (1989).

\bibitem{kotliar1986} G.~Kotliar and A.~E.~Ruckenstein, Phys.~Rev.~Lett. 
\textbf{57}, 1362 (1986).

\bibitem{gebhard1991} F.~Gebhard, Phys.~Rev.~B \textbf{44}, 992 (1991).

\bibitem{lechermann2007} F.~Lechermann, A.~Georges, G.~Kotliar, and
O.~Parcollet, Phys. Rev. B \textbf{76}, 155102 (2007).

\bibitem{bunemann2007} J.~B\"{u}nemann and F.~Gebhard, Phys. Rev. B \textbf{%
76}, 193104 (2007).

\bibitem{wang2008} G.-T.~Wang, X.~Dai, and Z.~Fang, Phys. Rev. Lett. 101,
066403 (2008).

\bibitem{wang2010} G.T.~Wang, Y.~Qian, G.~Xu, X.~Dai, and Z.~Fang, Phys.
Rev. Lett. 104, 047002 (2010).

\bibitem{tian2011} M.-F.~Tian, X.Y.~Deng, Z.~Fang, and X.~Dai, Phys. Rev. B
84, 205124 (2011).

\bibitem{KotliarCe} N. Lanat\`{a}, Y.--X. Yao, C.-Z. Wang, K.--M. Ho, J.
Schmalian, K. Haule, G. Kotliar, Phys. Rev. Lett. \textbf{111}, 196801
(2013).

\bibitem{FPLMTO} S. Y. Savrasov, Phys. Rev. B \textbf{54}, 16470 (1996).

\bibitem{onuki2012} Y.~Onuki and R.~Settai, Low Temp. Phys. \textbf{38}, 89
(2012).

\bibitem{LDA+U} For a review, see, e.g., V.~I.~ Anisimov, F.~Aryasetiawan,
and A.~I.~ Lichtenstein, J. Phys.: Condens. Matter \textbf{9}, 767 (1997).

\bibitem{Norman} C. S. Wang, M. R. Norman, R. C. Albers and A. M. Boring, W.
E. Pickett, H. Krakauer, N. E. Christensen, Phys. Rev. B 35, 7260 (1987).

%%%Ce

\bibitem{zolfl2001} M.~B.~Z\"olfl, I.~A.~Nekrasov, Th.~Pruschke,
V.~I.~Anisimov and J.~Keller, Phys. Rev. Lett. \textbf{87}, 276403 (2001).

\bibitem{held2001} K.~Held, A.~K.~McMahan and R.~T.~Scalettar, Phys. Rev.
Lett. \textbf{87}, 276404 (2001).

\bibitem{haule2005} K.~Haule, V.~Oudovenko, S.~Y.~Savrasov, and G.~Kotliar,
Phys. Rev. Lett. \textbf{94}, 036401 (2005).

\bibitem{amadon2006} B.~Amadon, S.~Biermann, A.~Georges, and
F.~Aryasetiawan, Phys. Rev. Lett. \textbf{96}, 066402 (2006).

\bibitem{PuNature} S. Y. Savrasov, G. Kotliar, E. Abrahams, Nature 410, 793
(2001).

\bibitem{vandereb2001} J.~W.~van der Eb, A.~B.~Kuz\~{O}menko, and D.~van der
Marel, Phys. Rev. Lett. \textbf{86}, 3407 (2001).

\bibitem{sakai2005} O.~Sakai, Y.~Shimizu and Y.~Kaneta, J. Phys. Soc. Jpn. 
\textbf{74}, 2517 (2005).

\bibitem{hall2001} D.~Hall, E.~C.~Palm, T.~P.~Murphy, S.~W.~Tozer, Z.~Fisk,
U.~Alver, R.~G.~Goodrich, J.~L.~Sarrao, P.~G.~Pagliuso, and T.~Ebihara,
Phys. Rev. B \textbf{64}, 212508 (2001).

\bibitem{settai2001} R.~Settai, H.~Shishido, S.~Ikeda, Y.~Murakawa,
M.~Nakashima, D.~Aoki, Y.~Haga, H.~Harima, and Y.~$\overline{\mathrm{O}}$%
nuki, J. Phys.: Condens. Matter \textbf{13}, L627 (2001).

\bibitem{elgazzar2004} S.~Elgazzar, I.~Opahle, R.~Hayn, and P.~M.~Oppeneer,
Phys. Rev. B \textbf{69}, 214510 (2004).

\bibitem{koitzsch2009} A.~Koitzsch, I.~Opahle, S.~Elgazzar, S.~V.~Borisenko,
J.~Geck, V.~B.~Zabolotnyy, D.~Inosov, H.~Shiozawa, M.~Richter, M.~Knupfer,
J.~Fink, B.~B\~{A}%
%TCIMACRO{\U{bc}}%
%BeginExpansion
$\frac14$%
%EndExpansion
chner, E.~D.~Bauer, J.~L.~Sarrao, and R.~Follath, Phys. Rev. B \textbf{79},
075104 (2009).

\bibitem{cornelius2000} A.~L.~Cornelius, A.~J.~Arko, J.~L.~Sarrao, and
N.~Harrison, Phys. Rev. B \textbf{62}, 14181 (2000).

\bibitem{hall2001_2} D.~Hall, E.~C.~Palm, T.~P.~Murphy, S.~W.~Tozer,
C.~Petrovic, E.~Miller-Ricci, L.~Peabody, C.~Q.~H.~Li, U.~Alver,
R.~G.~Goodrich, J.~L.~Sarrao, P.~G.~Pagliuso, J.~M.~Wills, and Z.~Fisk,
Phys. Rev. B \textbf{64}, 064506 (2001).

\bibitem{alver2001} U.~Alver, R.~G.~Goodrich, N.~Harrison, D.~W.~Hall, E.~C.
Palm, T.~P.~Murphy, S.~W.~Tozer, P.~G.~Pagliuso, N.~O.~Moreno, J.~L.~Sarrao,
and Z.~Fisk, Phys. Rev. B \textbf{64}, 180402 (R) (2001).

\bibitem{shishido2002} H.~Shishido, R.~Settai, D.~Aoki, S.~Ikeda,
H.~Nakawaki, N.~Nakamura, T.~Iizuka, Y.~Inada, K.~Sugiyama, T.~Takeuchi,
K.~Kindo, T.~C.~Kobayashi, Y.~Haga, H.~Harima, Y.~Aoki, T.~Namiki, H.~Sato,
and Y.~\={O}nuki, J. Phys. Soc. Jpn. \textbf{71}, 162 (2002).

\bibitem{shishido2005} H.~Shishido, R.~Settai, H.~Harima and Y.~\={O}nuki,
J. Phys. Soc. Jpn. \textbf{74}, 1103 (2005).

\bibitem{haga2001} Y.~Haga, Y.~Inada, H.~Harima, K.~Oikawa, M.~Murakawa,
H.~Nakawaki, Y.~Tokiwa, D.~Aoki, H.~Shishido, S.~Ikeda, N.~Watanabe, and Y.~%
\={O}nuki, Phys. Rev. B \textbf{63}, 060503(R) (2001).

\bibitem{haule2010} K.~Haule, C.~-H.~Yee, and K.~Kim, Phys. Rev. B \textbf{81%
}, 195107 (2010).

\bibitem{choi2012} H.~C.~Choi, B.~I.~Min, J.~H.~Shim, K.~Haule, and
G.~Kotliar, Phys. Rev. Lett. \textbf{108}, 016402 (2012)

\bibitem{fujimori2003} S.~I.~Fujimori, T.~Okane, J.~Okamoto, K.~Mamiya,
Y.~Muramatsu, A.~Fujimori, H.~Harima, D.~Aoki, S.~Ikeda, H.~Shishido,
Y.~Tokiwa, Y.~Haga, and Y.~$\overline{\mathrm{O}}$nuki, Phys. Rev. B \textbf{%
67}, 144507 (2003).

\bibitem{fujimori2006} S.~I.~Fujimori, A.~Fujimori, K.~Shimada, T.~Narimura,
K.~Kobayashi, H.~Namatame, M.~Taniguchi, H.~Harima, H.~Shishido, S.~Ikeda,
D.~Aoki, Y.~Tokiwa, Y.~Haga, and Y.~$\overline{\mathrm{O}}$nuki, Phys. Rev.
B \textbf{73}, 224517 (2006).

\bibitem{InterpolativeSolver} S. Y. Savrasov, K. Haule, V. Oudovenko, D.
Villani, G. Kotliar, Phys. Rev. B \textbf{71}, 115117 (2005).

\bibitem{szytula1981} A.~Szytula and I.~Szott, Solid State Commun. \textbf{40%
}, 199 (1981).

\bibitem{grier} B.~H.~Grier, J.~M.~Lawrence, V.~Murgai and R.~D.~Parks,
Phys. Rev. B \textbf{29}, 2664 (1984).

\bibitem{quezel1984} S.~Quezel, J.~Rossat-Mignod, B.~Chevalier, P.~Lejay and
J.~Etourneau, Solid State Commun. \textbf{49}, 685 (1984)

\bibitem{palstra} T.~T.~M.~Palstra, A.~A.~Menovsky, G.~J.~Nieuwenhuys and
J.~A.~Mydosh, J. Magn. Magn. Mater. \textbf{54-57}, 435 (1986).

\bibitem{movshovich1996} R.~Movshovich, T.~Graf, D.~Mandrus, J.~D.~Thompson,
J.~L.~Smith, and Z.~Fisk, Phys. Rev. B \textbf{53}, 8241 (1996).

\bibitem{mathur1998} N.~D.~Mathur, F.~M.~Grosche, S.~R.~Julian,
I.~R.~Walker, D.~M.~Freye, R.~K.~W.~Haselwimmer, and G.~G.~Lonzarich, Nature
(London) \textbf{394}, 39 (1998).

\bibitem{haen1987} P.~Haen, J.~Flouquet, F.~Lapierre, P.~Lejay, and
G.~Remenyi, J. Low Temp. Phys. \textbf{67}, 391 (1987).

\bibitem{aoki1992} H.~Aoki, S.~Uji, A.~K.~Albessard, and Y.~\={O}nuki, J.
Phys. Soc. Jpn. \textbf{61}, 3457 (1992).

\bibitem{yamagami1992} H.~Yamagami and A.~Hasegawa, J. Phys. Soc. Jpn. 
\textbf{61}, 2388 (1992).

\bibitem{yamagami1993} H.~Yamagami and A.~Hasegawa, J. Phys. Soc. Jpn. 
\textbf{62}, 592 (1993).

\bibitem{runge1995} E.~K.~R.~Runge, R.~C.~Albers, N.~E.~Christensen, and
G.~E.~Zwicknagl, Phys. Rev. B \textbf{51}, 10375 (1995).

\bibitem{onuki1992} Y.~\={O}nuki, I.~Umehara, A.~K.~Albessard, T.~Ebihara,
and K.~Satoh, J. Phys. Soc. Jpn. \textbf{61}, 960 (1992).

\bibitem{aoki1993} H.~Aoki, S.~Uji, A.~K.~Albessard, and Y.~\={O}nuki, J.
Phys. Soc. Jpn. \textbf{62}, 3157 (1993).

\bibitem{aoki1993prl} H.~Aoki, S.~Uji, A.~K.~Albessard, and Y.~\={O}nuki,
Phys. Rev. Lett. \textbf{71}, 2110 (1993).

\bibitem{endstra1993} T.~Endstra, G.~J.~Nieuwenhuys, and J.~A.~Mydosh, Phys.
Rev. B \textbf{48}, 9595 (1993).
\end{thebibliography}
\end{document}